\documentclass[aps,prl,showpacs,onecolumn,psfig,epsfig]{revtex4}

\draft
\usepackage{hyperref}
\usepackage{graphicx}
\newcommand{\Om}{\Omega}

\newcommand{\gm}{\gamma}

\newcommand{\om}{\omega}

\newcommand{\ve}{\varepsilon}
\newcommand{\pa}{\partial}

\begin{document}

\title{Surface solitons in left-handed metamaterials}
\author{G. T. Adamashvili$^{a,b,*}$ and A. Knorr${^b}$}
\affiliation{ Max-Planck-Institut  f\"ur Physik Komplexer
Systeme,N\"othnitzer Strasse 38, D-01187 Dresden, Germany $^{a}$\\
Institut f\"ur Theoretische Physik, Nichtlineare Optik und
Quantenelektronik, Technische Universit\"at  Berlin,
Hardenbergstr.36, D-10623 Berlin, Germany $^{b}$\\
$^*$ Permanent address: Georgian Technical University, Kostava str.
77, Tbilisi, 0179, Georgia. }

\begin{abstract}
 A theory of self-induced transparency of surface TM-mode propagating
along a interface separating conventional and left-handed
metamaterials is developed. A transition layer sandwiched between
connected media is described using a model of a two-dimensional gas
of quantum dots. Explicit analytical expressions for a surface
optical soliton in the presence of single and biexciton transitions,
depending on the magnetic permeability of the left-handed medium, are
obtained with realistic parameters which can be reached in current
experiments. It is shown that the sign of the total energy flow the
surface mode depends on the material parameters of the quantum dots and
the connected media.
\end{abstract}

%\Keywords{Metamaterial, Quantum dot, Optical soliton, Surface wave, Self-induced transparency}


\maketitle

The electromagnetic wave propagation through media is characterized by
the electric permittivity $\ve$, and the magnetic
permeability $\mu$, describing the coupling of the medium to
the electric and magnetic fields of the wave. The index of
refraction of the medium $n$ ($n^{2}=\ve \mu$) is determined by the
dispersion law for the electromagnetic plane wave. For plane waves
propagating  in isotropic regular media having simultaneously
positive  quantities $\ve$ and $\mu$, the vectors of the electric
$\vec{E}$  and  magnetic $\vec{H}$ fields and wave vector $\vec{k}$
constitute a  right-handed triplet of vectors. In such media the
Poynting vector $\vec{S}$, which is the cross product of the vector
of the electric and the magnetic fields,  does coincide with the wave
vector.  Such media are sometimes labeled as  the ordinary
right-handed materials (RHM). In conventional media the quantities
$\ve$ and $\mu$ are positive. But Veselago predicted that plane
electromagnetic wave propagation in an isotropic medium having
simultaneously negative permittivity  $\ve<0$ and permeability
$\mu<0$ have several other dramatically unusual electrodynamic
properties compared with regular RHM [1]. While the Poynting vector for
a plane wave still gives the direction of energy flow, the wave
vector will point in the opposite direction of $\vec{S}$ and vectors
$\vec{E},\;\;$ $\vec{H}$ and $\vec{k}$ shall form a left-handed
orthogonal set of vectors. In this case, they are termed left-handed
materials (LHM). Among unusual properties of the electromagnetic
wave  propagation in such media is negative refraction if the
wave refracted on the same side of the surface normal as the
incoming wave  at an RHM/LHM interface. Therefore the LHM is also
called a negative refractive index material (NIM). These artificial
metamaterials are fascinating because they exhibit properties that do not
occur in nature [2-4].

Pendry showed that a combination of split-ring resonators and
metallic wires can lead to a negative index of refraction [5]. The
main difficulty is to obtain a negative permeability $\mu<0$, which
does not occur in natural materials. Starting with first
realizations in the microwave region [6], recently, using silver as
a constituent material, the achieved demonstrate a negative index of
refraction at the red end of the visible spectrum (780 nm wavelength
)[7].

Negative refraction could occur for surface electromagnetic waves
(SEW) propagation in layered structures. Surface modes can
propagate in a frequency region where the permittivities of two
connected media have the opposite signs, i.e. SEW propagate at the
interface between media with different "rightness". Lately SEW have
attracted much interest in nano-optics [8], diverse
applications [9] and also can be used as a convenient tool for
studying properties of the structured surfaces considered
as optical metamaterials [10]. Characterized peculiarity of these
waves are strong enhancement and spatial confinement of the
electromagnetic field of the wave near to the interface. In NIM
interest to  SEW  is connected with the strong impact of the SEW on
the image resolution of an LHM flat lens [11]. The properties of the
SEW propagating at the interface between conventional and LHM  as
well as between two different LHM
have been investigated for a
intensity-dependent dielectric
permittivity [12-14].

Another type resonant nonlinear waves can be formed within the
McCall-Hahn mechanism, where a nonlinear coherent interaction takes
place via Rabi-oscillations of the carrier density, if the
conditions for self-induced transparency (SIT) are fulfilled:
\begin{equation}
\omega T>>1,\;\;\;\; T<<T_{1,2},
\end{equation}
in attenuator medium. The steady-state $2\pi $ -pulse (soliton) is
generated, when $\Theta >\pi$ [15]. Here $T$ and $\omega $ are the
width and the carrier frequency of the pulse, $T_{1}$ and $T_{2}$
are the longitudinal and transverse relaxation times of the atomic
systems, $\Theta$  is the area of the pulse. In conventional media
SIT for SEW  for atomic systems and semiconductor quantum dots (SQD)
have been investigated [16-18]. SQD, also referred to as
zero-dimensional atoms(artificial atoms), are nanostructures which
allow confinement of the charge carriers in all three spatial
directions, resulting in atomic-like discrete energy spectra with
strongly enhanced carrier lifetimes[19]. Such features make quantum
dots in many respects similar to atoms. SIT in SQD have been
investigated experimentally [20] and theoretically [21,22] for plane
waves and waveguide modes.

The purpose of the present Letter is to theoretically investigate
the processes of the formation of surface optical solitons under SIT
conditions in a resonance transition layer of two-dimensional
sheet of inhomogeneously broadened SQD sandwiched at
the LHM/RHM interface.

We consider the formation of an optical soliton for surface TM-modes
in a SQD transition layer sandwiched between conventional RHM and
the LHM interface in the case when a optical pulse with width $T$ and
frequency $\omega>>T^{-1}$ is propagating along positive direction
of the $z$ axis. On the flat border of the division (x=0) between
two isotropic media, a thin transition layer with thickness $d$
containing a small concentration of SQD having the polarization
$\vec P(x,z,t)=\vec e_{p}p(z,t)\delta(x)$, where $\vec e_{p}$ is the
polarization unit vector along $z$ axis. Semi-spaces are divided at
$x>0$ and $x<0$, with conventional RHM (medium I) and LHM (medium
II) with different electric permittivities $\ve_{1}(\om)>0$ and
$\ve_{2}(\om)<0$ and the magnetic permeabilities  $\mu_{1}(\om)>0$
and $\mu_{2}(\om)<0$ , respectively. In optical region of spectra
$d<<\lambda$, where $\lambda$ is length of the surface optical wave
[19]. Therefore, SIT can be modeled as SEW propagation along the
interface LHM/RHM and infinite small thickness transition layer
(monolayer) of SQDs.

For a surface TM-mode the electric field $\vec{E}(E_{x},0,E_{z})$
lies in the $xz$ plane perpendicular to the boundary of the division
between two connected media and the magnetic field
$\vec{H}(0,H_{y},0)$ is directed along the axis $y$. The quantity
\begin{equation}
U_{1}(x,z,t)=\int U_{1}(\Om,Q)e^{\kappa_{1}x}e^{i(Qz-\Om t)} d\Om
dQ,\;\;\;\;\;\;\;for\;\;\;\;x<0,
$$$$
U_{2}(x,z,t)=\int U_{2}(\Om,Q)e^{-\kappa_{2}x}e^{i(Qz-\Om t)} d\Om
dQ,\;\;\;\;\;\;\;for\;\;\;\;x>0,
\end{equation}
where
\begin{equation}
  \kappa_{i}^{2}=Q^{2}- \ve_{i}(\Om)\mu_{i}(\Om)
 \frac{\Om^{2} }{c^{2}},\;\;\;\;\;\;\;i=1,2.
\end{equation}or a Fourier-decomposition of the fields.
The functions $U_{1,2}$ stand for the components
$(E_{x},E_{z},H_{y})$ in both connected media. We assume
translational invariance in the $y$-direction, so that all field
quantities  do not depend from the coordinate $y$, i.e. $\frac{\pa
}{\pa y} \rightarrow 0.$

Taking into account the surface current caused by presence of the
SQDs, the boundary conditions for surface waves at $x=0$ read
[16,18,23]:
\begin{equation}
H_{2,y}-H_{1,y}= \frac{4\pi}{c} \frac{\pa p}{\pa
t},\;\;\;\;\;\;{E}_{1,z}={E}_{2,z}.
\end{equation}
Using the eqs.(2)-(4)we obtain the nonlinear wave equation for the
$E_{z}$ component of the strength of electrical field at $x=0$, in
the following form[18]:
\begin{equation}
\int f(\Om,Q)E(\Om,Q)e^{i(Qz-\Om t)} d\Om dQ=-4\pi p(z,t),
\end{equation}
where
\begin{equation}
 f(\Om,Q)=\frac{\ve_{2} }{ \kappa_{2}} + \frac{\ve_{1}}{\kappa_{1}},
 \end{equation}
$E_{1,z}(\Om,Q)=E_{2,z}(\Om,Q)=E(\Om,Q)$. This equation is valid for
any dependence of the polarization of the SQD $p(z,t)$ on the
strength of electrical field at $x=0$. In order to determine the
dependence of the polarization $p(z,t)$ on the strength of
electrical field at $x=0$, we have to consider the structure of the
energetic levels of the SQD and the details of the nonlinear
interaction of the surface pulse with the SQD. We assume that the
SQD can be described by the ground $|0>$, exciton $|2>$ and
biexciton $|3>$ states.

The Hamiltonian of the system [18,21,22]
$$
H=H_0 +V,
$$
where
$$
H_{0}=\hbar\omega_{12}|2><2|+\hbar\omega_{13}|3><3|
$$
is the Hamiltonian of the single-exciton and biexciton states and
$V=-\vec P  \vec E$ is the Hamiltonian of the light-quantum dots
interaction, $\hbar$ is the Planck's constant, $\om_{ij}$ are
frequencies of excitations between energetic levels $i$ and $j$
$\;(i,j,=1,2,3)$.

In general, the detunings from the resonance $\omega_{13} -
\omega_{12} - \omega$ and $\omega_{12} - \omega$, which describe the
SQDs, are different. Under the assumption of off-resonant excitation
with a constant detuning
$\omega_{13}-\omega_{12}-\omega\approx\omega_{12}-\omega=\Delta$ the
polarization which is determined by interband transitions occurring
in the quantum dots between the three energetic levels
\begin{equation}
p=N \int g(\Delta )[{\mu}_{12}\rho_{21}(\Delta)+{\mu}_{23}
\rho_{32}(\Delta) ] d \Delta +c.c.,
\end{equation}
where $N$ is the uniform quantum dot density,
 ${\mu}_{12}=\vec{\mu}_{12} \vec e,\;\;\;{\mu}_{23}=\vec{\mu}_{23}
\vec e; \;\;\vec{\mu}_{12}$ and $ \vec{\mu}_{23}$ are the dipole
elements for the corresponding transitions, $\vec{e}$ is the
polarization unite vector along $\vec{E}$. We assume the dipole
moments to be parallel to each other and to be directed along axis
$z$,$\;{\mu}_{12}={\mu}_{23}$; $g(\Delta \omega)$ is the
inhomogeneous broadening lineshape function resulting from dots size
fluctuations. The quantities $\rho_{ij}$ are the matrix elements of
the density matrix $\rho$ are determined by the Liouville equation
$$i\hbar\frac{\partial {\rho}_{nm}}{\partial
t}=\sum_{l}(<n|H|l>{\rho}_{lm}-{\rho}_{nl}<l|H|m>),
$$
where $n,m,l,=1,2,3$.

The solution of this equation we can find following the way
presented in the work [18]. We can simplify Eq. (5) using the method
of slowly changing profiles. For this purpose, we represent the
functions $E$ and $p$ in the form
\begin{equation}
 E=\sum_{l=\pm1}\hat{E}_{l} Z_l;\;\;\;\;p=\mu_{12} N \tilde{p}\; Z_{1}+ c.c.
\end{equation}
where $\hat{E}_{l}$ and $\tilde{p}$  are the slowly varying complex
amplitudes of the optical electric field and the polarization,
$Z_{l}= e^{il(kz -\omega t)}$. To guarantee that $E$ is a real
number, we set $\hat{E}_{l}=\hat{E}_{-l}^{\ast}=\hat{E}$. This
approximation is based on the consideration that the envelopes
$\hat{E}$ vary sufficiently slowly in space and time as compared
with the carrier wave parts-i.e.,
\begin{equation}
|\frac{\partial \hat{E}}{\partial t}|<<\omega
|\hat{E}|,\;\;\;|\frac{\partial \hat{E}}{\partial z }|<<k|\hat{E}|,
\end{equation}
and is called the slowly varying envelope approximation [15].

Substituting the equations (6) and (8) in the wave equation (5), and
to take into account the explicit form of the envelope of the
polarization (7) which determined from the Liouville equation, after
divided the real and imaginary parts of the equation (5) we obtain
dispersion law for surface pulse propagating on the interface
between left-handed and right-handed media
\begin{equation}
k^{2} = \frac{\om^{2}}{ c^{2}}
 \frac{ \ve_{2}\ve_{1}}{\ve_{2} +\ve_{1}}
\frac{ \ve_{2}\mu_{1} - \ve_{1}\mu_{2}}{\ve_{2} -\ve_{1}}
\end{equation}
and a nonlinear wave equation in the form:
\begin{equation}\label{}
 (\frac{d \hat{E}}{d\zeta})^2 =T^{-2} {\hat{E}}^2  -\frac{{\mu_{12}}^2}{2{\hbar}^2 }
   {\hat{E}}^4,
\end{equation}
where  the width of the pulse $T$ is determined by the equation
\begin{equation}\label{}
T^{-2}=\frac{4\pi N {\mu}^{2}_{12}}{ f'_{\Om}(\frac{v_g}{V}-1)
\hbar}\int g(\Delta')F (\Delta')d \Delta',
\end{equation}
$c$ is the speed of light in vacuum, $\zeta=t-\frac{z}{V}\;$, $V$ is
the constant pulse velocity.
\begin{equation}
v_{g}=-\frac{f'_{Q}}{f'_{\Om}}, \;\;\;\;\;\;\;\;\;\;\;\;\;\; f'_{Q}=
\frac{\pa f}{\pa Q}|_{\Om=\om,Q=k}=-k[
\frac{\ve_{1}(\om)}{\tilde{\kappa}^{3}_{1} } +
\frac{\ve_{2}(\om)}{\tilde{\kappa}^{3}_{2} }],
$$$$
f'_{\Om}=\frac{\pa f}{\pa \Om}|_{\Om=\om,Q=k}=\sum_{i=1,2}
\frac{1}{\kappa_{i}}\{ \frac{d \ve_{i}}{d \Om} +
\frac{\ve_{i}}{\kappa^{2}_{i}}\;\; \frac{ \Om}{2c^{2}}\;\; [2
\ve_{i}\mu_{i}+\Om \frac{d (\ve_{i}\mu_{i})}{d \Om}
]\}|_{\Om=\om,Q=k},
\end{equation}
where
\begin{equation}
\tilde{\kappa}_{i}^{2}(\om, k)=k^{2}-\ve_{i}(\om)\mu_{i}(\om)
\frac{\om^{2}}{c^{2}},
\end{equation}
$\om$ and $k$ are frequency and wave number of the carrier wave, $
v_{g}$ is group velocity of linear SEW.

The solution of Eq.(11) for the envelope function has the form
[15,18]
\begin{equation}
\hat {E}=\frac{2}{\mu_{0}T} sech{  \frac{t-\frac{z}{V}}{T}},
\end{equation}
where $\mu_{0}=\sqrt{2}\frac{{\mu_{12}}}{{\hbar} }$ and we can
determine the constant velocity of the $2\pi$ pulse of the SEW:
\begin{equation}\label{}
\frac{1}{V}=\frac{1}{v_g}+\frac{4\pi N {\mu}^{2}_{12}}{\hbar
f'_{\Om} v_g }\int \frac{ g(\Delta') d
\Delta'}{T^{-2}+{{\Delta}'}^{2}}.
\end{equation}

In optical frequency region the magnetic permeability
loses its usual physical meaning, in RHM we must put magnetic
permeability $\mu_{1}=1$, otherwise would be an over-refinement[23].
We have to note that the description of an isotropic NIM in terms
$\ve(\om)$ and $\mu(\om)$ is not unique. Besides the $\ve(\om)$ and
$\mu(\om)$ approach for NIM, there is also an alternative
description,  based on the generalized, spatial dispersive
permittivity $\tilde{\ve}(\om,\vec{k})$. In this approach, the
non-local function $\tilde{\ve}(\om,\vec{k})$, depending besides
$\om$ also on the wave vector $\vec{k}$, describes both electrical
and magnetic responses of the medium and different optical effects
in NIM [24].

The Eqs. (2),(10),(12)-(16) determine the parameters of the surface
soliton for any value of $x,z$ and $t$ and show that for the
existence of a soliton it is necessary that conditions $f'_{\Om}>0$
and $V < v_g$ are fulfilled. Parameters of the surface optical
solitons depend not only on the SQD parameters and the permittivites
of the two interface media, but (unlike the case of interface of the
two conventional connected media [18] ) also depends on the magnetic
permeability $\mu_{2}(\om)$ and its derivative
$\frac{d\mu_{2}(\Om)}{d\Om}|_{\Om=\om}$ of the LHM.

The time-averaged over a period $2\pi/\om$ of oscillation of the
field Poynting vector of the TM mode  $\langle \vec{S} \rangle$,
which is associated with the energy flow of the pulse, has the $z$
components in the RHM-medium 1 and LHM- medium 2 in the following
form[13,25,26]
\begin{equation}
\langle S_{1,z}\rangle =\frac{c^{2}k}{8\pi\om
\ve_{1}}|\hat{H}_{1}|^{2}e^{2
\kappa_{1}x},\;\;\;\;\;\;\;\;\;\;\;\;\;\;\;\;\;\;\;\; \langle
S_{2,z}\rangle =\frac{c^{2}k}{8\pi\om \ve_{2}}|\hat{H}_{2}|^{2}e^{-2
\kappa_{2}x}
\end{equation}
The corresponding total energy flow
\begin{equation}
N=\int_{-\infty}^{0}\langle {S_{1,z}}\rangle
dx+\int_{0}^{+\infty}\langle {S_{2,z}}\rangle
dx=\frac{c^{2}k}{16\pi\om}(\frac{|\hat{H}_{1}|^{2}}{\ve_{1}\kappa_{1}}+
\frac{|\hat{H}_{2}|^{2}}{\ve_{2}\kappa_{2}}).
\end{equation}
To take into account that in LHM $\ve_{2}=-|\ve_{2}|$ the quantities
$\langle S_{1,z} \rangle$ and $\langle S_{2,z} \rangle$ have
opposite signs and consequently the surface TM-mode resonance
soliton will be have a vortexlike distribution of the Poynting
vector [13,25].

Equation (18) has the general form and valid for any boundary
condition for the strength of magnetic field of the SEW. The total
energy flow is positive and coincide with the wave vector when the
condition is satisfied:
\begin{equation}
\kappa_{1}  \ve_{1}|\hat{H}_{2}|^{2}<\kappa_{2} |\ve_{2}|
|\hat{H}_{1}|^{2}.
\end{equation}
In case when the transition layer is absent, the boundary condition
for envelopes of the strength of magnetic field of the surface
TM-mode has the following form: $ \hat{H}_{2}=\hat{H}_{1} $, and
because for SEW $\ve_{1}<|\ve_{2}|$ and $\kappa_{1}<\kappa_{2}$, the
condition (19) always is satisfied and total energy flow $N$ is
always positive [25,26].

The situation is changed when the resonance transition layer is
included. From the boundary condition (4) the connection between $
\hat{H}_{2}$ and $\hat{H}_{1} $ has following form
\begin{equation}
|\hat{H}_{2}|^{2}= |\hat{H}_{1}|^{2}  + i R(\tilde{p}^{*}
\hat{H}_{1}- \tilde{p} \hat{H}^{*}_{1}) +R^{2}| \tilde{p} |^{2}
\end{equation}
where
$$
R=\frac{4\pi \om \mu_{12} N}{c}.
$$
From equation (20) it is evident that the resonance transition layer
cause the change the total energy flow of the surface TM-mode and it
will be positive or negative depending from the parameter $R$ and
polarization $\tilde{p}$. When the condition
\begin{equation}
\kappa_{1} \ve_{1} [|\hat{H}_{1}|^{2}  + i R(\tilde{p}^{*}
\hat{H}_{1}- \tilde{p} \hat{H}^{*}_{1}) +R^{2}| \tilde{p}
|^{2}]<\kappa_{2} |\ve_{2}|   |\hat{H}_{1}|^{2}
\end{equation}
is fulfilled the total energy flow will be positive (forward SEW),
otherwise negative (backward SEW).

In the considered case, the quantities $ \hat{H}_{2}$ and
$\hat{H}_{1} $ are real functions and hence the equations (20) is
simplified
\begin{equation}
\kappa_{1} \ve_{1} [\hat{H}^{2}_{1}  + i R(\tilde{p}^{*} - \tilde{p}
)\hat{H}_{1} +R^{2}| \tilde{p} |^{2}]< \kappa_{2} |\ve_{2}|
\hat{H}^{2}_{1}.
\end{equation}

Consequently, when the resonance transition layer is included,
depending from the parameters of the SEW, SQD and the connected LHM/RHM,
the total energy flow can have a positive or negative sign as
well. This property is general and are valid for any  transition
layer which causes the surface polarization (current) and
consequently change the boundary conditions. This statement are
valid for any linear or nonlinear wave processes for TM-mode in NIM
and also for SEW propagating along interface between conventional
media. Under the condition $N=0$ we obtain the localization  conditions of the surface excitations.

In the optical NIM precious metallic nanostructures are used,
therefore the losses are significant for the surface optical waves in
LHM. But in present work the questions regarding of the losses of
the surface optical modes and their amplifications are not
considered (see, for example [27,28]). In metals and LHM the SEW
occurs when the carrier frequency $\om$ is below of plasma frequency
$\om_{p}$. For the coherent interaction of the pulse of surface wave
with medium the duration of the pulse  should be much shorter than
the characteristic plasmonic oscillation damping time $2\pi/ \gm$.
Silver is known as significantly lower losses than other metals at
the optical frequencies. The plasma frequency for silver $\om_{p}=2\pi
\times 2,18 \cdot 10^{15}\;s^{-1}$and damping rate is $\gamma=2\pi
\times 5,08 \cdot 10^{12}\;s^{-1}$ [3]. On the other hand a planar
array of split-ring resonators can be fabricated upon gallium
arsenide (GaAs) substrate [29]. On the surface or at the boundary of
semiconductors (for instant GaAs or InAs) with another medium, a
small concentration of SQD can be grown. Soliton of the SEW in
regular media when one of the connected medium was semiconductor
with SQD is investigated [18]. SQD is one of the promising object
also for amplification of the optical waves [28,30]. Transverse
relaxation times of the quantum dots, which are of order of
nanoseconds to several tens of picoseconds [31,32], is longer than
the characteristic plasmonic oscillation damping time  $2\pi/ \gm$.
The envelope approximation (9) is appropriate for pulses when the
spectral width of the pulse much smaller than the $\om$ and valid
for pulses with width $T\gtrsim 20 fs$ [33]. Consequently pulse with
such width which acceptable for investigation of the nonresonance
nonlinear waves in the LHM [12,13], is satisfying also the eq.(1)
the conditions of SIT  in SQD . This circumstances allow to hope
that SIT in LHM/RHM with SQD transition layer can be experimentally
observable.

In conclusion, we considered case when
a transition layer contains SQD, however, the presented
results have a  more general character and are valid also in case when
the transition layer contains usual optical impurity atoms sandwiched
between LHM/RHM.

Acknowledgements

A.G.T.would like to thank G. Stegeman, V. Shalaev and M. Wegener
for discussions, the Max-Planck-Institut  f\"ur Physik Komplexer
Systeme and Technical University of Berlin (Deutsche
Forschungsgemeinschaft) for financial support and hospitality, ISTC
under grant G-1219 for financial support.

\end{document}